\documentstyle[psfig,amssymb]{oldllncs}
\begin{document}

\newcommand{\beq}{\begin{equation}}
\newcommand{\eeq}{\end{equation}}
\newcommand{\bea}{\begin{eqnarray*}}
\newcommand{\eea}{\end{eqnarray*}}
\newcommand{\eps}{\epsilon}
\newcommand{\ord}{{\cal O}}
\newcommand{\Z}{{\Bbb Z}}
\newcommand{\cho}{\left( \! \begin{array}{c}}
\newcommand{\ose}{\end{array} \! \right)}
\newcommand{\mat}{\scriptsize \left( \! \begin{array}{cc}}
\newcommand{\maat}{\scriptsize \left( \! \begin{array}{ccc}}
\newcommand{\Mat}{\scriptsize \left( \! \begin{array}{cccc}}
\newcommand{\rix}{\end{array} \! \right)}
\newcommand{\PP}{{\bf P}}
\newcommand{\BP}{{\bf BP}}
\newcommand{\QBP}{{\bf QBP}}
\newcommand{\QP}{{\bf QP}}
\newcommand{\NC}{{\bf NC}}
\newcommand{\QNC}{{\bf QNC}}
\newcommand{\bra}{\langle}
\newcommand{\ket}{\rangle}
\newcommand{\R}{{\Bbb R}}

\newtheorem{conjecture}[theorem]{Conjecture}{\bfseries}{\itshape}

\title{Some Notes on Parallel Quantum Computation}
\author{Cristopher Moore\inst{1} and Martin Nilsson\inst{2}}
\institute{Santa Fe Institute, 1399 Hyde Park Road,
Santa Fe, New Mexico 87501 {\tt moore@santafe.edu} \and
Chalmers Tekniska H\"ogskola and University of G\"oteborg, G\"oteborg, Sweden
{\tt martin@fy.chalmers.se}}
\maketitle

\begin{abstract}
We exhibit some simple gadgets useful in designing shallow parallel circuits
for quantum algorithms.  We prove that any quantum circuit composed entirely 
of controlled-not gates or of diagonal gates can be parallelized to 
logarithmic depth, while circuits composed of both cannot.  Finally, 
while we note the Quantum Fourier Transform can be parallelized to
linear depth, we exhibit a simple quantum circuit related to it
that we believe cannot be parallelized to less than linear depth, 
and therefore might be used to prove that $\QNC < \QP$.
\end{abstract}

Much of computational complexity theory has focused on the question
of what problems can be solved in polynomial time.  Shor's quantum
factoring algorithm \cite{shor} suggests that quantum computers might be 
more powerful than classical computers in this regard, i.e.\ that
$\QBP$ might be a larger class than $\PP$, or rather $\BP$, the class
of problems solvable in polynomial time by a classical probabilistic 
Turing machine with bounded error.

More recently, a distinction has been made between $\PP$ and the class
$\NC$ of efficient parallel computation, namely the subset of $\PP$ of 
problems which can be solved by a parallel computer with a polynomial
number of processors in {\em polylogarithmic} time, i.e.\ $\ord(\log^k n)$
time for some $k$, where $n$ is the number of bits of the input \cite{papa}.
Equivalently, $\NC$ problems are those solvable by Boolean circuits
with a polynomial number of gates and polylogarithmic depth.

This distinction seems especially relevant for quantum computers,
where decoherence makes it difficult to do more than a limited number
of computation steps reliably.  Since decoherence due to storage errors
is essentially a function of time, we can avoid it by doing as many 
of our quantum operations at once as possible; if we can parallelize our 
computation to logarithmic depth, we can solve exponentially larger problems.
(Gate errors, on the other hand, will not be improved by parallelization,
and may even get worse if the parallel algorithm involves more gates.)

We define quantum operators and quantum circuits as follows:

\begin{definition}  A {\em quantum operator} on $n$ qubits is a
unitary rank-$2n$ tensor $U$ where 
$U_{a_1 a_2 \ldots a_n}^{b_1 b_2 \ldots b_n}$
is the amplitude of the incoming and outgoing truth values being
$a_1, a_2, \ldots a_n$ and $b_1, b_2, \ldots b_n$ respectively,
with $a_i, b_i \in \{0,1\}$ for all $i$.  However, we will usually 
write $U$ as a $2^n \times 2^n$ unitary matrix $U_{ab}$ where 
$a$ and $b$'s binary representations are $a_1 a_2 \cdots a_n$ and
$b_1 b_2 \cdots b_n$ respectively.

A {\em one-layer circuit} consists of the tensor product of one- and 
two-qubit gates, i.e.\ rank 2 and 4 tensors, or $2 \times 2$ and $4 \times 4$ 
unitary matrices.  This is an operator that can be carried out by a set of 
simultaneous one-qubit and two-qubit gates, where each qubit interacts with 
at most one gate.

A {\em quantum circuit of depth $k$} is a quantum operator written as the 
product of $k$ one-layer circuits.  The {\em depth} of a quantum operator 
is the depth of the shallowest circuit equal to it.
\end{definition}

Here we are allowing arbitrary two-qubit gates.  If we like, we can restrict
this to {\em controlled-$U$ gates}, of the form
$\Mat 1 & 0 & 0 & 0 \\
      0 & 1 & 0 & 0 \\
      0 & 0 & u_{11} & u_{12} \\
      0 & 0 & u_{21} & u_{22} \rix$, or more stringently to the 
{\em controlled-not} gate
$\Mat 1 & 0 & 0 & 0 \\
      0 & 1 & 0 & 0 \\
      0 & 0 & 0 & 1 \\
      0 & 0 & 1 & 0 \rix$.  For these, we will call the first and second
qubits the {\em input} and {\em target} qubit respectively, even though
they don't really leave the input qubit unchanged, since they entangle it
with the target.

Since either of these can be combined with one-qubit gates to simulate 
arbitrary two-qubit gates \cite{barenco}, these restrictions would just 
multiply our definition of depth by a constant.  The same is true if we 
wish to allow gates that couple $k > 2$ qubits as long as $k$ is fixed, 
since any $k$-qubit gate can be simulated by some constant number of 
two-qubit gates.

\begin{figure}
\centerline{\psfig{file=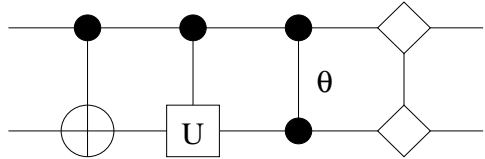,width=2.5in}}
\caption{Our notation for controlled-not, controlled-$U$,
symmetric phase shift, and arbitrary diagonal gates.}
\end{figure}

In order to design a shallow parallel circuit for a given quantum operator,
we want to be able to use additional qubits or ``ancillae'' for intermediate 
steps in the computation, equivalent to additional processors in a parallel 
quantum computer.  However, to avoid entanglement, we demand that our 
ancillae start and end in a pure state $|0\ket$, so that the desired 
operator appears as the diagonal block of the operator performed by the 
circuit on the subspace where the ancillae are zero.

Then in analogy with $\NC$ we propose the following definition:

\begin{definition}  Let $F$ be a family of quantum operators,
i.e.\ $F(n)$ is a $2^n \times 2^n$ unitary matrix on $n$ qubits.
We say that $F(n)$ is {\em embedded} in an operator $M$ with $m$ ancillae 
if $M$ is a $2^{m+n} \times 2^{m+n}$ matrix which preserves 
the subspace where the ancillae are set to $|0\ket$, and if $M$ is 
identical to $F(n) \otimes \bf{1}^{2^m}$ when restricted to this subspace.

Then $F$ is in $\QNC^k$ if, for some constants $c_1$, $c_2$ and $j$,
$F(n)$ can be embedded in a circuit of depth at most $c_1 \log^k n$, 
with at most $c_2 n^j$ ancillae.  Then $\QNC = \cup_k \QNC^k$, the 
set of operators parallelizable to polylogarithmic depth with a 
polynomial number of ancillae.
\end{definition}

To extend this definition from quantum operators to decision problems in the 
classical sense, we have to choose a measurement protocol, and to what
extent we want errors bounded.  We will not explore those issues here.

We will use the notation in figure 1 for our various gates:
the controlled-not and controlled-$U$, the symmetric phase shift
$\Mat 1 & 0 & 0 & 0 \\
      0 & 1 & 0 & 0 \\
      0 & 0 & 1 & 0 \\
      0 & 0 & 0 & e^{i\theta} \rix$, and arbitrary diagonal gates
$\Mat e^{i\theta_{00}} & 0 & 0 & 0 \\
      0 & e^{i\theta_{01}} & 0 & 0 \\
      0 & 0 & e^{i\theta_{10}} & 0 \\
      0 & 0 & 0 & e^{i\theta_{11}} \rix$.

\section{Permutations}

In classical circuits, one can move wires around as much as one likes.
In a quantum computer, it may be more difficult to move a qubit from
place to place.  However, we can easily show that we can do arbitrary
permutations in constant depth:

\begin{proposition}  Any permutation of $n$ qubits can be performed
in 4 layers of controlled-not gates with $n$ ancillae, or in 6 layers
with no ancillae.
\label{permprop}
\end{proposition}

\begin{figure}
\centerline{\psfig{file=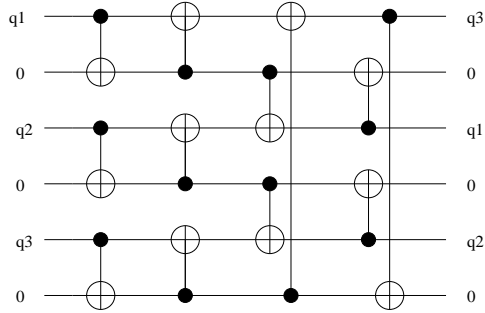,width=2.5in}}
\caption{Permuting $n$ qubits in $4$ layers using $n$ ancillae.}
\label{switch2}
\end{figure}

\begin{proof}
The first part is obvious; simply copy the qubits into the ancillae,
cancel the originals, recopy them from the ancillae in the desired order,
and cancel the ancillae.  This is shown in figure \ref{switch2}.

\begin{figure}
\centerline{\psfig{file=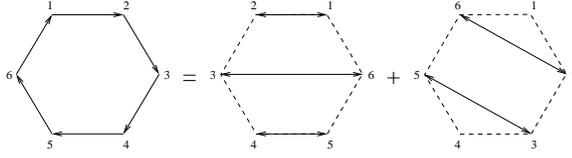,width=3in}}
\caption{Any cycle, and therefore any permutation, is the
composition of two sets of disjoint transpositions.}
\label{permutation}
\end{figure}

\begin{figure}
\centerline{\psfig{file=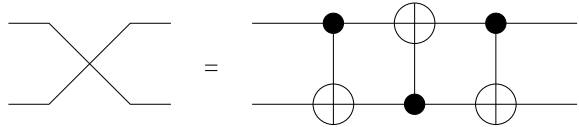,width=3in}}
\caption{Switching two qubits with three controlled-nots.}
\label{switch}
\end{figure}

Without ancillae, we can use the fact that any permutation can be written 
as the composition of two sets of disjoint transpositions \cite{perm}.  
To see this, first decompose it into a product of disjoint cycles, 
and then note that a cycle is the composition of two reflections, 
as shown in figure \ref{permutation}.  Two qubits can be switched 
with 3 layers of controlled-not gates as shown in figure \ref{switch}, 
so any permutation can be done in 6 layers.
\qed \end{proof}

\section{Fan-out}

To make a shallow parallel circuit, it is often important to {\em fan out} 
one of the inputs into multiple copies.  The controlled-not gate can be used 
to copy a qubit onto an ancilla in the pure state $|0\ket$ by making a 
non-destructive measurement: 
\[ (\alpha |0\ket + \beta |1\ket ) \otimes |0\ket \,\to\, 
   \alpha |00\ket + \beta |11\ket \]
Note that the final state is not a tensor product of two independent qubits;
the two qubits are completely entangled.  Making an unentangled copy requires 
non-unitary, and in fact non-linear, processes since 
\[ (\alpha |0\ket + \beta |1\ket) \otimes (\alpha |0\ket + \beta |1\ket) 
   = \alpha^2 |00\ket + \alpha \beta (|01\ket + |10\ket) + \beta^2 |11\ket \]
has coefficients quadratic in $\alpha$ and $\beta$.

This means that disentangling or {\em uncopying} the ancillae by the 
end of the computation, returning them to their initial state $|0\ket$, 
is a non-trivial and important part of a quantum circuit.  There are, 
however, some special cases where this can be done easily.    

Suppose we have a series of $n$ controlled-$U$ gates all with the same
input qubit.  Rather than applying them in series, we can {\em fan out} 
the input into $n$ copies by splitting it $\log_2 n$ times, apply them
to the target qubits, and uncopy them afterward, thus reducing the
circuit's depth to $\ord(\log n)$ depth.

\begin{figure}
\centerline{\psfig{file=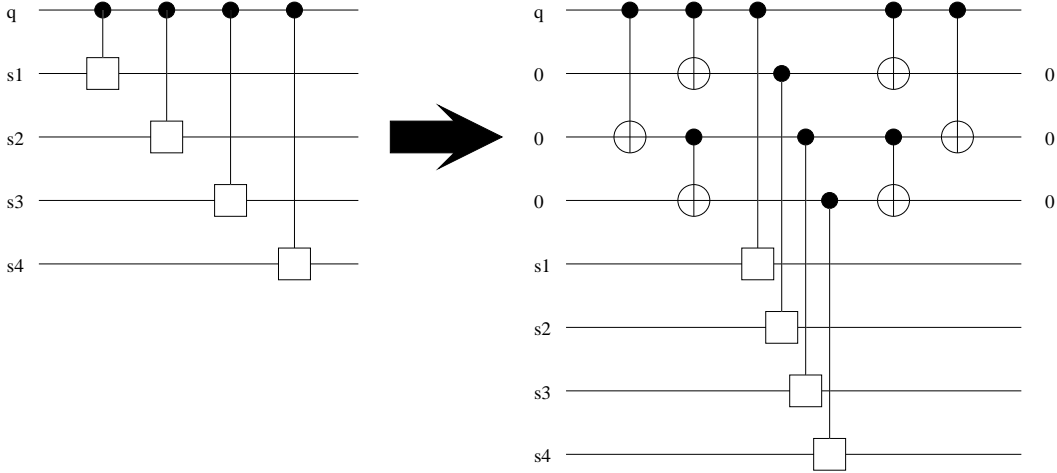,width=5.5in}}
\caption{Parallelizing $n$ controlled gates on a single input qubit $q$
to $\ord(\log n)$ depth.}
\label{fanoutfig}
\end{figure}

\begin{proposition}  A series of $n$ controlled gates coupling the same input 
to $n$ target qubits can be parallelized to $\ord(\log n)$ depth with 
$\ord(n)$ ancillae.  
\label{fanoutprop}
\end{proposition}

\begin{proof}
The circuit in figure \ref{fanoutfig} copies the input onto $n-1$ ancillae, 
applies all the controlled gates simultaneously, and uncopies the ancillae 
back to their original state.  Its total depth is $2 \log_2 n + 1$.
\qed \end{proof}

This kind of symmetric circuit, in which we uncopy the ancillae to 
return them to their original state, is similar to circuits 
designed by the Reversible Computation Group at MIT \cite{knight} 
for reversible classical computers.

\section{Diagonal and mutually commuting gates}

Fan-in seems more difficult in general.  Classically, if a single
qubit receives controlled gates from $n$ inputs, we can calculate 
the composition of these in $\ord(\log n)$ time by composing them
in pairs, but it is unclear when we can do this with unitary 
linear operators.  One special case where it is possible is if
all the gates are diagonal:

\begin{figure}
\centerline{\psfig{file=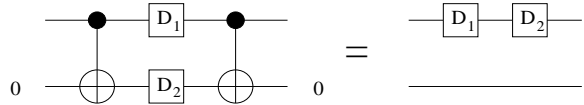,width=3in}}
\caption{Using entanglement to parallelize diagonal operators.}
\label{entanglefig}
\end{figure}

\begin{figure}
\centerline{\psfig{file=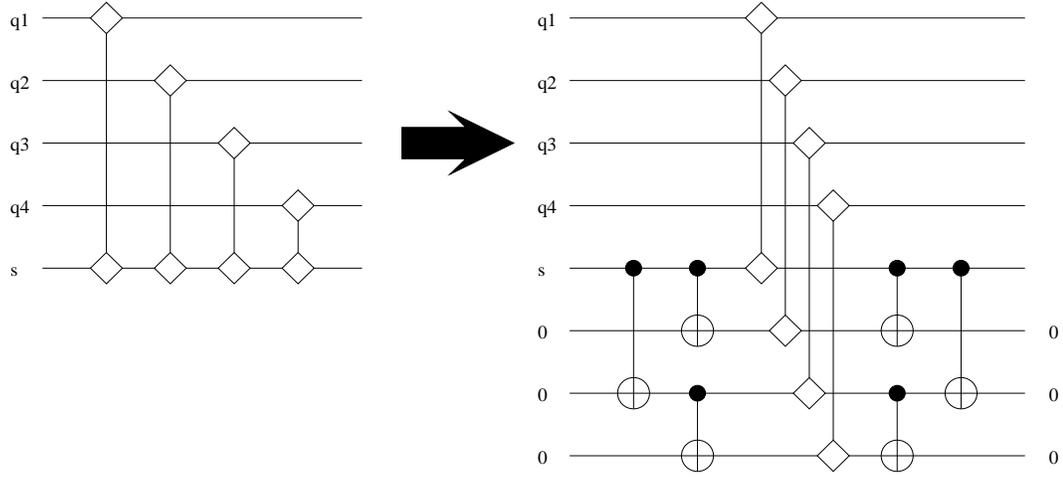,width=5.5in}}
\caption{Parallelizing $n$ diagonal gates on a single qubit as in 
proposition \ref{fanoutprop}.}
\label{diagfig}
\end{figure}

\begin{proposition}
A series of $n$ diagonal gates coupling the same qubit to $n$ others
can be parallelized to $\ord(\log n)$ depth with $\ord(n)$ ancillae.
\label{diagprop}
\end{proposition}

\begin{proof}
Here the entanglement between two copies of a qubit becomes an asset.
Since diagonal matrices don't mix Boolean states with each other,
we can act on a qubit and an entangled copy of it with two diagonal matrices
$D_1$ and $D_2$ as in figure \ref{entanglefig}.  When we uncopy the
ancilla, we have the same effect as if we had applied both matrices
to the original.  Then the same kind of circuit as in proposition 
\ref{fanoutprop} works, as shown in figure \ref{diagfig}.
\qed \end{proof}

Since matrices commute if and only if they can be simultaneously
diagonalized, we can generalize this to the case where a set of 
controlled-$U$ gates applied to a given target qubit have 
mutually commuting $U$s:

\begin{proposition}
A series of of $n$ controlled-$U$ gates acting on a single qubit,
where the $U$s mutually commute, can be parallelized to $\ord(\log n)$ depth 
with $\ord(n)$ ancillae.
\label{commuteprop}
\end{proposition}

\begin{proof}
Since the $U$s all commute, they can all be diagonalized by the same
unitary operator $T$.  Apply $T$ to the target qubit, parallelize the 
circuit using proposition \ref{diagprop}, and put the target qubit 
back in the original basis by applying $T^{-1}$.  This is all done 
with a circuit of depth $2 \log_2 n + 3$. 
\qed \end{proof}

\begin{figure}
\centerline{\psfig{file=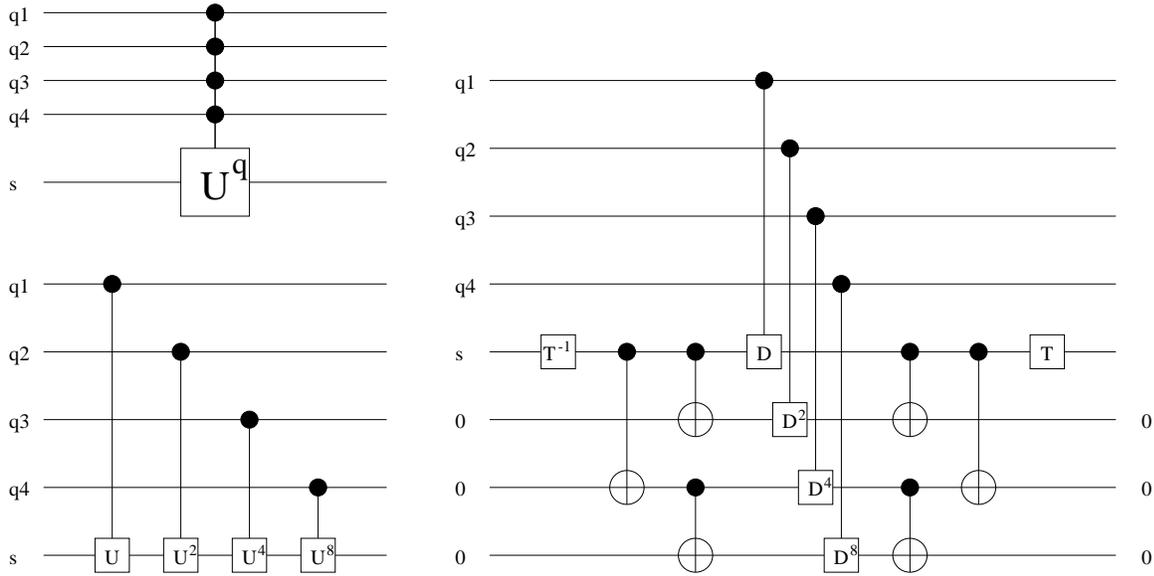,width=6in}}
\caption{Applying an operator $U$ $q$ times, where $q$ is given in binary 
by the control qubits.}
\label{powerfig}
\end{figure}

As an example, in figure \ref{powerfig} we show a circuit that applies
the $q$th power of an operator $U$ to a target qubit, where
$0 \le q < 2^k$ is given by $k$ control qubits as a binary integer. 
We can do this because $U, U^2, U^4, \ldots$ can be simultaneously 
diagonalized, since $U^q = T D^q T^{-1}$.  Note that this works
for operators $U$ that act on any number of qubits.

We can extend this to circuits in general whose gates are mutually
commuting, which includes diagonal gates:

\begin{proposition}
Any circuit consisting of diagonal or mutually commuting gates, 
each of which couples at most $k$ qubits, can be parallelized to 
depth $\ord(n^{k-1})$ with no ancillae, and to depth $\ord(\log n)$ 
with $\ord(n^k)$ ancillae.  Therefore, any family of such circuits
is in $\QNC^1$.
\label{diagqnc}
\end{proposition}

\begin{proof}
Since all the gates commute, we can sort them by which qubits 
they couple, and arrive at a compressed circuit with one gate for each 
$k$-tuple.  This gives $\cho n \\ k \ose = \ord(n^k)$ gates,
but by performing groups of $n/k$ disjoint gates simultaneously 
we can do all of them in depth $\ord(n^{k-1})$.

By making $\ord(n^{k-1})$ copies of each qubit, we can then use
propositions \ref{diagprop} and \ref{commuteprop} to reduce this
further to $\ord(\log n)$ depth.  
\qed \end{proof}

This is hardly surprising; after all, diagonal gates are just conditional
phase shifts, and saying that two gates commute is almost like saying
that they can be performed simultaneously.

\section{Circuits of controlled-not gates}

We can also fan in controlled-not gates.  Figure \ref{fig4} shows how to 
implement $n$ controlled-not gates on the same target qubit in depth 
$2 \log_2 n+1$.  The ancillae carry the intermediate ``sums mod 2'' 
of the inputs, and we add them in pairs.

\begin{figure}
\centerline{\psfig{file=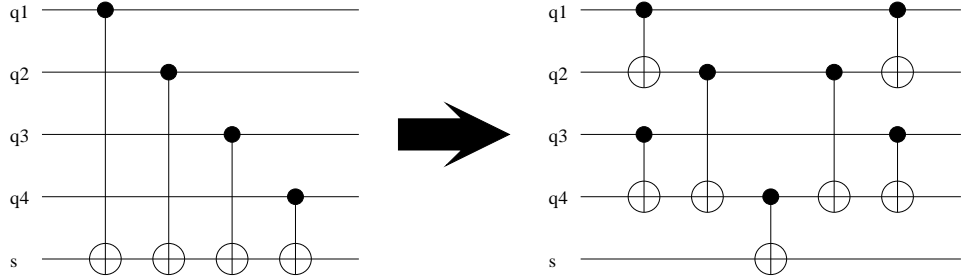,width=5in}}
\caption{Parallelizing $n$ controlled-not gates to $\ord(\log n)$ depth
by adding them in pairs.}
\label{fig4}
\end{figure}

We can use a generalization of this circuit to show that any circuit
composed entirely of controlled-not gates can be parallelized to
logarithmic depth:

\begin{proposition}  Any circuit on $n$ qubits composed entirely of 
controlled-not gates can be parallelized to $\ord(\log n)$ depth 
with $\ord(n^2)$ ancillae.  Therefore, any family of such circuits
is in $\QNC^1$.
\label{xorqnc}
\end{proposition}

\begin{proof}
First, note that in any circuit of controlled-not gates, if the $n$ 
input qubits have binary values and are given by an $n$-dimensional vector 
$q$, then the output can be written $Mq$ where $M$ is an $n \times n$
matrix over the integers mod 2.  Each of the output qubits can be written 
as a sum of up to $n$ inputs, $(Mq)_i = \sum_k q_{j_k}$ where $j_k$ 
are those $j$ for which $M_{ij}=1$.

We can break these sums down into binary trees.  Let $W_n$ be the
complete output sums, $W_{n/2}$ be their left and right halves consisting 
of up to $n/2$ inputs, and so on down to single inputs.  There are less than 
$n^2$ such intermediate sums $W_k$ with $k > 1$.  We assign an ancilla to 
each one, and build them up from the inputs in $\log_2 n$ stages,
adding pairs from $W_k$ to make $W_{2k}$.  The first stage takes 
$\ord(\log n)$ time and an additional $\ord(n^2)$ ancillae since 
we may need to add each input into multiple members of $W_2$, 
but each stage after that can be done in depth 2.

\begin{figure}
\centerline{\psfig{file=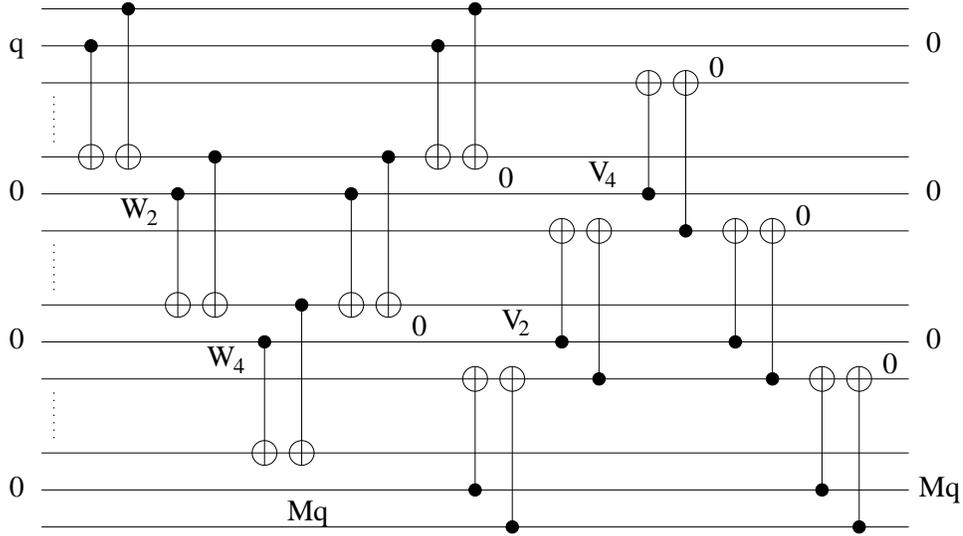,width=5in}}
\caption{Parallelizing an arbitrary circuit of controlled-not gates
to logarithmic depth.}
\label{xorfig}
\end{figure}

To cancel the ancillae, we use the same cascade in reverse order, 
adding pairs from $W_k$ to cancel $W_{2k}$.  This leaves us 
with the input $q$, the output $Mq$, and the ancillae set to zero.

Now we use the fact that, since the circuit is unitary, $M$ is invertible.
Thus we can recalculate the input $q=M^{-1}(Mq)$ and cancel it.  
We use the same ancillae in reverse order, building the inputs $q$ 
out of $Mq$ with a series of partial sums $V_2, V_4, \ldots$, 
cancel $q$, and cancel the ancillae in reverse as before.
All this is illustrated in figure \ref{xorfig}.

This leaves us with the output $Mq$ and all other qubits zero.
With four more layers as in proposition \ref{permprop}, we can shift 
the output back to the input qubits, and we're done.
\qed \end{proof}

This result is hardly surprising; after all, these circuits are 
reversible Boolean circuits, and any classical circuit composed of 
controlled-not gates is in $\NC^1$.  We just did a little
extra work to disentangle the ancillae.

\section{Circuits with both diagonal and controlled-not gates}

We have shown that circuits composed of diagonal or controlled-not gates
can be parallelized.  Since circuits composed of both these kinds of gates 
have only one non-zero element in each row and column, they are really just 
classical reversible circuits with phase shifts attached.  Therefore,
it's reasonable to ask whether propositions \ref{diagqnc} and \ref{xorqnc}
can be combined; that is, whether arbitrary circuits composed of phase shifts
and controlled-not gates can be parallelized to logarithmic depth.

In this section, we will show that this is not the case.
However, this will not help us show that $\QNC < \QP$.

\begin{proposition}  Any diagonal unitary operator on $n$ qubits can be 
performed by a circuit consisting of an exponential number of controlled-not 
gates and one-qubit diagonal gates and no ancillae.
\label{morseprop}
\end{proposition}

\begin{proof}
Any diagonal unitary operator on $n$ qubits consists of $2^n$ phase shifts, 
$\maat e^{i\omega_0} & & \\ & \ddots & \\ & & e^{i\omega_{2^n-1}} \rix$.
If we write the phase angles as a $2^n$-dimensional vector $\omega$, 
then the effect of composing two diagonal operators is simply to 
add these vectors mod $2\pi$.

For each subset $s$ of the set of qubits, define a vector
$\mu_s$ as $+1$ if the number of true qubits in $s$ is even, 
and $-1$ if it is odd.  If $s$ is all the qubits, for instance,
$\mu_{\{1\ldots n\}}$ is the aperiodic Morse sequence
$(+1,-1,-1,+1,\ldots)$ when written out linearly, but it really just means 
giving the odd and even nodes of the Boolean $n$-cube opposite signs.  

\begin{figure}
\centerline{\psfig{file=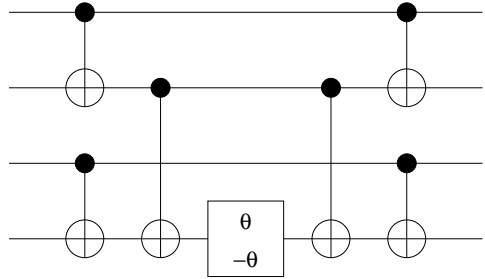,width=2.5in}}
\caption{A circuit for the phase shift $\theta \mu_s$, i.e.\ a phase shift
of $+\theta$ if the number of true qubits is even and $-\theta$ if it is odd.}
\label{morsefig}
\end{figure}

It is easy to see that the $\mu_s$ for all $s \subset \{1,\ldots,n\}$
are linearly independent, and form a basis of $\R^{2^n}$.
Moreover, while diagonal gates coupling $k$ qubits can only perform
phase shifts spanned by those $\mu_s$ with $|s| \le k$, the circuit 
in figure \ref{morsefig} can perform a phase shift proportional to $\mu_s$ 
for any $s$ (incidentally, in depth $\ord(\log |S|)$ with no ancillae).  
Therefore, a series of $2^n$ such circuits, one for each subset of 
$\{1,\ldots,n\}$, can express any diagonal unitary operator.
\qed \end{proof}

Then we have the following corollary:

\begin{corollary}  There are circuits composed of controlled-not gates
and one-qubit diagonal gates that cannot be parallelized to less than
exponential depth with a polynomial number of ancillae.
\end{corollary}

\begin{proof}  Consider setting up a many-to-one correspondence
between circuits and operators.  The set of diagonal unitary operators 
on $n$ qubits has $2^n$ continuous degrees of freedom, while the set of 
circuits of depth $d$ and $m$ ancillae has only $\ord(d(m+n))$ continuous 
degrees of freedom (and some discrete ones for the circuit's topology).  
Thus if $m$ is polynomial, $d$ must be exponential.
\qed \end{proof}

Note that this counting argument does not help us distinguish $\QP$ 
from $\QNC$, since both have a polynomial number of degrees of freedom.  
Neither does it help us exhibit a particular family of circuits which 
require exponential depth, since it is completely non-constructive.  
The classical situation is similar; there are $2^{2^n}$ Boolean functions 
on $n$ variables, but only $2^{\ord(dw \log w)}$ circuits of depth $d$ 
and width $w$.  Thus the vast majority of Boolean functions require 
exponential depth if the width is polynomial, but proving a lower bound
on the depth of a particular one remains elusive.

\section{$\QNC \ne \QP$?  The staircase circuit}

\begin{figure}
\centerline{\psfig{file=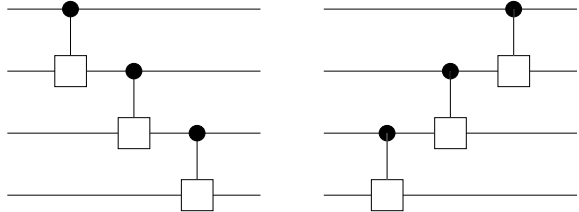,width=3in}}
\caption{This ``staircase'' circuit seems hard to parallelize 
unless the operators are purely diagonal or off-diagonal.}
\label{staircasefig}
\end{figure}

A simple, perhaps minimal, example of a quantum circuit that seems 
hard to parallelize is the ``staircase'' circuit shown in figure 
\ref{staircasefig}.  This kind of structure appears in the standard circuit 
for the quantum Fourier transform, which has $\ord(n^2)$ gates
\cite{coppersmith,shor}.  Careful inspection shows that the QFT 
can in fact be parallelized to $\ord(n)$ depth  as shown in figure 
\ref{qftfig} \cite{griffiths}, but it seems difficult to do any better.
Clearly, any fast parallel circuit for the QFT would be relevant to 
prime factoring and other problems the QFT is used for.  

If we define $\QP$ as the family of quantum operators that can be expressed 
with circuits of polynomial depth (again, leaving measurement issues aside 
for now), we can make the following conjecture:

\begin{conjecture}  Staircase circuits composed of controlled-$U$ gates
other than diagonal or off-diagonal gates (i.e.\ other than the special
cases handled in propositions \ref{diagqnc} and \ref{xorqnc}) 
cannot be parallelized to less than linear depth.  Therefore, $\QNC < \QP$.
\end{conjecture}

\begin{figure}
\centerline{\psfig{file=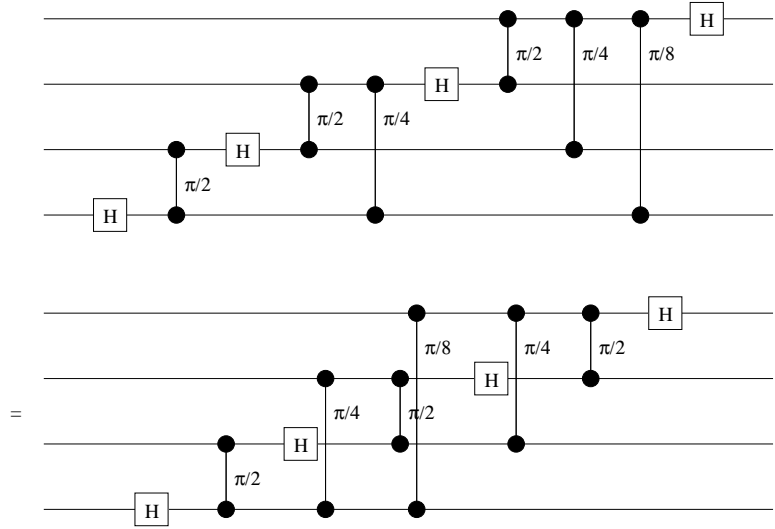,width=4in}}
\caption{The standard circuit for the quantum Fourier transform on 
$n$ qubits can be carried out in $2n - 1$ layers.  Can we do better?}
\label{qftfig}
\end{figure}

\section{Conclusion}

We conclude with some questions for further work.

Parsing classical context-free languages is in $\NC$.  Quantum context-free 
languages have been defined in \cite{qrl}.  Is quantum parsing, i.e.\ producing
derivation trees with the appropriate amplitudes, in $\QNC$?

Can circuits for quantum error correction such as those in \cite{ds}
be parallelized to significantly smaller depth?  If so, does this reduce 
the threshold error necessary for long-term computation, at least as far as 
storage errors are concerned?

Finally, can the reader show that the staircase circuit cannot be
parallelized, thus showing that $\QNC < \QP$?  This would be 
quite significant, since corresponding classical question $\NC < \PP$
is still open.

{\bf Acknowledgments.}  M.N. would like to thank the Santa Fe Institute 
for their hospitality, and Spootie the Cat for her affections.  C.M. would 
like to thank the organizers of the First International Conference 
on Unconventional Models of Computation in Auckland, New Zealand, 
as well as Seth Lloyd, Tom Knight, David DiVincenzo and Artur Ekert for 
helpful conversations.  He would also like to thank Molly Rose for 
inspiration and companionship.  This work was supported in part by 
NSF grant ASC-9503162.

\end{document}